\documentclass{aa}
\usepackage{graphicx}
\usepackage{natbib}
\usepackage{hyperref}
\usepackage{txfonts}
\begin{document}

   \title{The far reaches of the $\beta$ Pictoris debris disk
\thanks{Based on archival observations from the European Southern Observatory, Chile (Program 072.C-0299).}
}


   \author{Markus Janson\inst{1} \and
               Alexis Brandeker\inst{1} \and
               G{\"o}ran Olofsson\inst{1} \and
               Ren{\'e} Liseau\inst{2}
          }

   \institute{Department of Astronomy, Stockholm University, Stockholm, Sweden\\
              \email{markus.janson@astro.su.se}
        \and
            Department of Space, Earth and Environment, Chalmers University of Technology, Onsala Space Observatory, Onsala, Sweden
             }

   \date{Received ---; accepted ---}

   \abstract{The nearby young star $\beta$~Pictoris hosts a rich and complex planetary system, with at least two giant planets and a nearly edge-on debris disk that contains several dynamical subpopulations of planetesimals. While the inner ranges of the debris disk have been studied extensively, less information is known about the outer, fainter parts of the disk. Here we present an analysis of archival FORS V-band imaging data from 2003-2004, which have previously not been explored scientifically because the halo substructure of the bright stellar point spread function is complex. Through a high-contrast scheme based on angular differential imaging, with a forward-modelling approach to mitigate self-subtraction, we produced the deepest imaging yet for the outer range of the $\beta$~Pic disk, and extracted its morphological characteristics. A  brightness asymmetry between the two arms of the edge-on disk, which was previously noted in the inner disk, is even more pronounced at larger angular separations, reaching a factor $\sim$10 around 1000\,AU. Approaching 2000\,AU, the brighter arm is visible at a surface brightness of 27--28 mag/arcsec$^2$. Much like for the brightness asymmetry, a tilt angle asymmetry exists between the two arms that becomes increasingly extreme at large separations. The outer tilt angle of 7.2 deg can only be explained if the outer disk is farther from an edge-on inclination than the inner disk, or if its dust has a stronger scattering anisotropy, or (most likely) both. The strong asymmetries imply the presence of a highly eccentric kinematic disk component, which may have been caused by a disruptive event thought to have taken place at a closer-in location in the disk.}


\titlerunning{Far reaches of the $\beta$ Pic debris disk}
\authorrunning{M. Janson et al.}

   \maketitle
%

\section{Introduction}
\label{s:intro}

$\beta$~Pictoris (also known as $\beta$~Pic, HD 39060, HIP 27321, or HR 2020) is an A6V-type star hosting both a disk and a planetary system. It is one of the most frequently studied such systems in the past several decades. The debris disk around the star was originally detected in the form of infrared excess by the IRAS satellite \citep{aumann1985}, and shortly thereafter became the first circumstellar disk to be spatially resolved \citep{smith1984}. In addition to being nearby at 19.44--19.75 pc \citep{vanleeuwen2007,brown2018} and bright with a magnitude of $V=3.86$ mag of the primary star, the $\beta$~Pic disk is seen largely edge-on, with high resulting column densities along the line of sight, all of which benefits detectability relative to other disk host systems. On the celestial sphere, the disk extends between the north-east (NE) and south-west (SW) directions, and we correspondingly refer to the two arms of the disk as the NE and the SW arms.

The disk has been studied intensively both in terms of its dust \citep[e.g.][]{weinberger2003,liseau2003,okamoto2004} and in terms of its gas \citep[e.g.][]{slettebak1975,dent2014,cavallius2019}. The dust reveals fundamental morphological features and population substructures within the disk. For example, the NE arm of the disk is known to be substantially brighter in dust continuum than the SW arm \citep{kalas1995}, which might imply eccentric components within the disk. Moreover, a warp has been observed in the disk \citep{heap2000} that could be characterised as a secondary disk component superimposed on the primary one \citep{golimowski2006}. This could be interpreted as revealing the existence of a giant planet on a slightly inclined orbit with respect to the main disk plane, dynamically inducing the secondary disk \citep{mouillet1997}. In 2009 and later, this was confirmed when planet $\beta$~Pic b was discovered \citep{lagrange2009,lagrange2010}, which was found to be on an orbit compatible with creating the warp \citep{chauvin2012}. The planet has since been redetected and characterised through several different methods \citep[e.g.][]{snellen2014,snellen2018,hoeijmakers2018,nowak2020}, and a second planet ($\beta$~Pic c) has recently been announced \citep{lagrange2019}, further underlining the richness of the $\beta$~Pic disk and planetary system.

Gas in the disk has been observed both through transmission spectroscopy of material (in some cases, comets) passing in front of the star \citep[e.g.][]{vidal1994,roberge2006,kiefer2014}, and through emission spectroscopy farther out in the disk \citep[e.g.][]{dent2014,brandeker2016,cataldi2018}. In the SW arm of the disk, CO and CI gas have been observed to be concentrated in a clump between 50 and 100\,AU projected separation. Because of the dissociation timescales in the high-energy radiative circumstellar environment of $\beta$~Pic, much of the localised gas is thought to have been released very recently. This supports a scenario in which a large planetesimal was disrupted in the relatively recent past and had its fragments scattered over a range of eccentric orbits \citep{jackson2014}. Once per orbit, all of the fragments converge back on the collisional point, causing frequent collisions, and ensuring a fresh supply of gas released in the process, thus potentially explaining the observed properties of the gas. While the hypothesised disruption appears to have taken place at a separation in the 50--100\,AU range, the eccentric distribution of the aftermath will largely reside at much larger separations. Further clues to explain this event might therefore be found in the outer parts of the disk.

As indicated above, the $\beta$~Pic planet and disk system has been studied in considerable detail, but as a result of limitations in the field of view (and/or contrast and sensitivity), such studies have typically been limited to separations within a few hundred AU from the parent star. A wider-field coronagraphic image presented in \citet{larwood2001} has revealed that the disk extends much farther than this, although the relatively modest telescope size (2.2 m) limited the sensitivity for characterizing the outermost parts of the disk. In late 2003 and early 2004, a deep-imaging sequence was launched with the FORS1 camera at the 8.2 m European Southern Observatories (ESO) Very Large Telescope (VLT), in program 072.C-0299 with R.\ Liseau as principal investigator. The data were of very good average quality, but difficulties associated with the outer regions of the bright stellar point spread function (PSF) prevented an accurate analysis of the disk at the time, which led to the data set being shelved. Since that time, considerable progress has been made in the field of high-contrast processing and PSF modeling and subtraction. Meanwhile, the FORS data set from 2003-2004 remains the deepest imaging sequence for $\beta$~Pic ever acquired (as far as we can identify), and the best chance for characterizing the outer ($\sim$500--2000\,AU) regions of the disk. Thus, we were motivated to revisit this archival data set in the context of modern PSF subtraction techniques and analyse the PSF-free outer disk. This paper is the outcome of that effort. 

The paper is organised as follows: We outline the properties of the archival data and the initial steps of data reduction in Sect. \ref{s:obs}. In Sect. \ref{s:psf}, we describe the PSF subtraction schemes tested and the final procedure for generating an image that is (to first order) free from the impact of the stellar PSF, with a particular emphasis on mitigating self-subtraction of the disk flux  in Sect. \ref{s:selfsub}. The results of this procedure and corresponding analysis are presented in Sect. \ref{s:analysis}. We then discuss some implications of these results in Sect. \ref{s:discussion}, and finally summarise our conclusions in Sect. \ref{s:summary}.

\section{Observations and data reduction}
\label{s:obs}

All observations for this programme were acquired in the $V$ band with the FORS1 camera at the ESO VLT (the `Antu' unit telescope) between 28 November 2003 and 23 February 2004. The program,e was executed in nine observing sessions, but because one had poor ambient conditions, eight data sets are included in the analysis. Each data set contains five image frames of the target $\beta$~Pic and five frames of a reference star, $\alpha$~Pic. $\alpha$~Pic was chosen as a reference star because it happens to be similar to $\beta$~Pic in many respects: The brightness and spectral type of $\beta$~Pic are $V=3.86$ mag and A6V, while the corresponding quantities for $\alpha$~Pic are $V = 3.30$ mag and A8VnkA6. Being separated by about 1h in right ascension, the two stars are also placed well in the sky to be observed back-to-back within an observational set. Each $\beta$~Pic frame consisted of a single 250 s exposure, and each $\alpha$~Pic frame consisted of a 154 s exposure (matching the integration times after the relative brightnesses). As a result, across the 40 frames per object, the total effective on-source exposure time was 2.8 hours for $\beta$~Pic and 1.7 hours for $\alpha$~Pic. With overheads and including both targets, the total VLT time spent on the programme was approximately 7.5 hours. This was a partial completion because the initially approved allocation was 13 hours.

The high-resolution imaging mode of FORS1 was used, which corresponds to a pixel scale of 0.09975 arcsec/pixel, and thus a square field of view $\sim$200$^{\prime \prime}$ across. For all astrometric values quoted in this paper, we have taken the mild distortion of the FORS1 field into account using the equation for this purpose given in the FORS manual. During the observation, an opaque bar of width $\sim$47$^{\prime \prime}$ was placed on top of the bright central star (for both $\alpha$~Pic and $\beta$~Pic) in order to block out as much of the stellar PSF as possible and to simplify detection of faint material in the circumstellar environment. The bar was placed horizontally in the field and was fixed with respect to the sky (not with respect to the telescope pupil), which creates challenges in the data-processing step, as we discuss in Section \ref{s:psf}. 

The fundamental data reduction steps were provided by the ESO FORS pipeline, which includes steps such as bias- and dark subtraction and flat-field corrections. The remaining procedure was performed with a custom reduction procedure. While the star was always placed at approximately the same position during the observations, there were small drifts between frames and small offsets between observation sets, so that in order to register and recentre the frames to a common frame of reference, we used the bright PSF spider arms sticking out from below the mask. By determining centroids along each arm and making linear fits to the results, we determined the stellar photocentre of each frame as the location in which the linear fits to the two sets of arms overlapped. The relative position angle of the x-pattern formed by the spider arms between different frames also gave a high-quality estimate of the relative parallactic angles from frame to frame. The parallactic angle variation within each individual session was 5\degr--9\degr, while the variation between sessions was in excess of 120\degr. These angles were later used in the PSF subtraction scheme described in the next section. To prepare the data for this PSF subtraction, the frames were shifted to a common centre using spline interpolation. 

\section{PSF subtraction}
\label{s:psf}

\subsection{Subtraction strategy}

\begin{figure*}[htb]
\centering
\includegraphics[width=16cm]{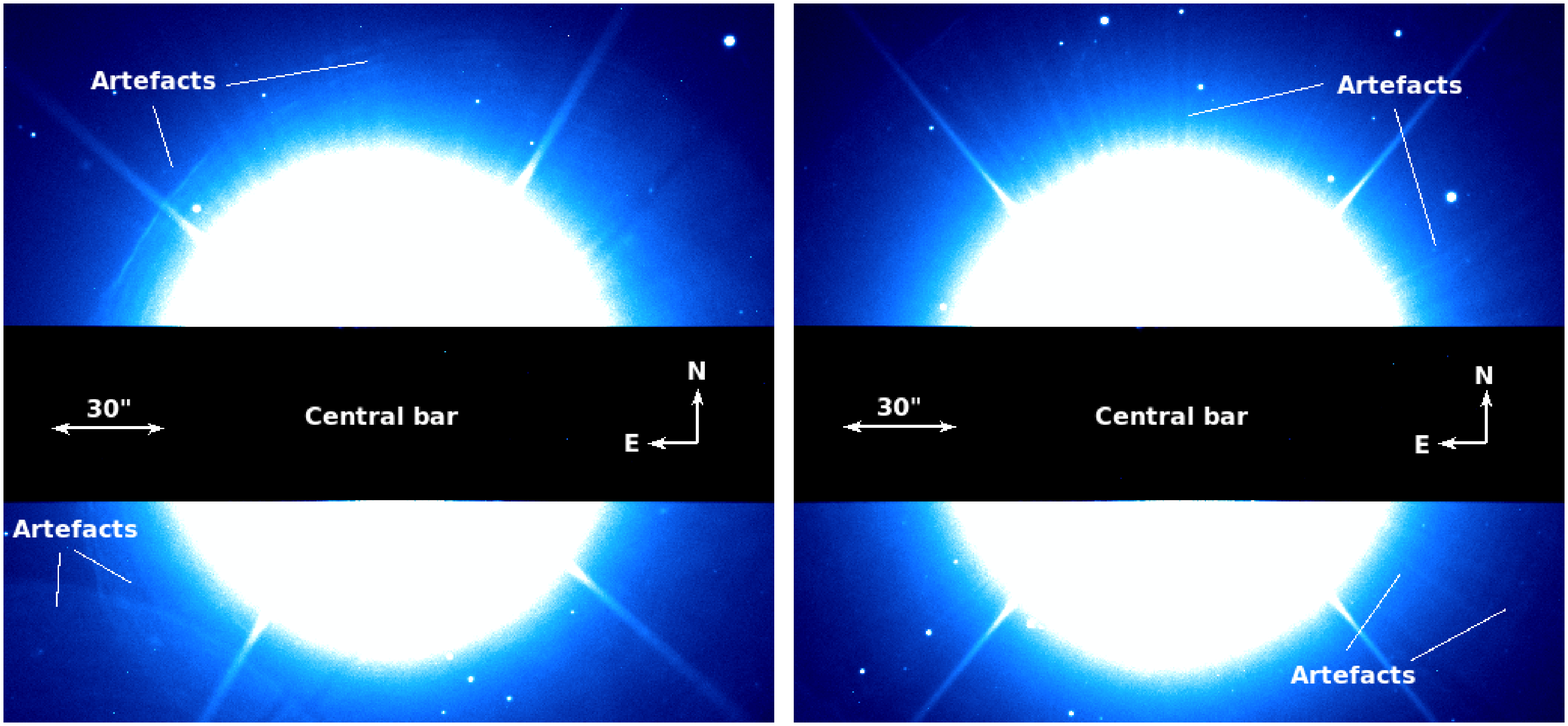}
\caption{Images of $\beta$~Pic (left) and the PSF reference star $\alpha$~Pic (right) at an image stretch focusing on the faint outer parts of the PSF. The two frames are as close together in time as possible with only the slew from one target to the next in between. Significant differences between the two PSF representations still persist, however, making it challenging to use $\alpha$~Pic as reference for $\beta$~Pic.}
\label{f:bothartefacts}
\end{figure*}

\begin{figure}[htb]
\centering
\includegraphics[width=8cm]{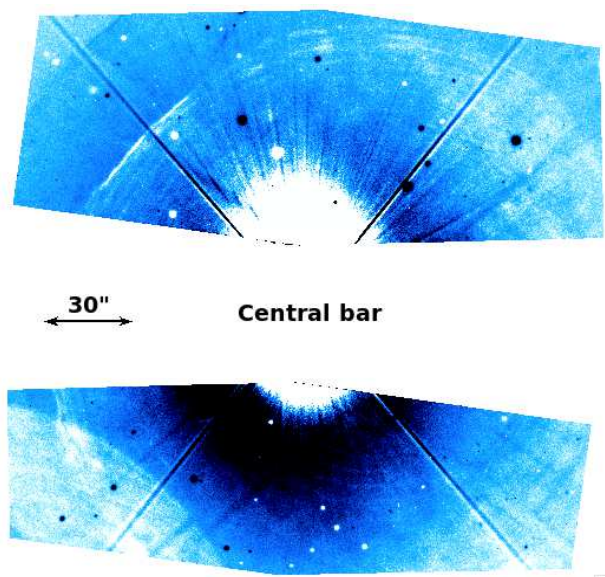}
\caption{Same as Fig. \ref{f:bothartefacts}, but here the two images have been rotated to a mutual pupil angle and one has been subtracted from the other. This allows us to distinguish the systematic PSF differences more clearly, which include asymmetrically distributed PSF halos, different sharpness for the spider arms (probably due to different degrees of pupil smearing between $\alpha$ Pic and $\beta$ Pic in the field-stabilised observations), and most critically for our purposes, a range of artefacts in the outer parts of the image domain, which are not in common in the $\beta$~Pic and $\alpha$~Pic images.}
\label{f:betaminusalpha}
\end{figure}

The $\alpha$~Pic observations acquired adjacently to each $\beta$~Pic observation could in principle have been almost ideal for PSF referencing because the two stars are similar in brightness and spectral type and because the observations were obtained relatively closely on the sky and in time. The PSF subtraction scheme would then simply have consisted of finding ideal matching between representations of the $\alpha$~Pic PSF and the $\beta$~Pic PSF, for example by pairing representations that match closely in time, or by constructing an idealised PSF model for each $\beta$~Pic frame based on principal component analysis (PCA) using the $\alpha$~Pic frames as a training set \cite[e.g.][]{soummer2012,amara2012}. We tried several such strategies, but unfortunately, it turns out that $\alpha$~Pic simply does not function as a suitable PSF reference for $\beta$~Pic to a sufficient level of precision so that a high final image depth would be reached. As Fig. \ref{f:bothartefacts} shows, a pair of $\alpha$~Pic and $\beta$~Pic frames can display very different PSF substructure patterns, even if they are taken relatively closely together in time. An alternative display of these differences is shown in Fig. \ref{f:betaminusalpha}. This is probably not due to changing seeing or similar, but rather due to scattered or reflected light in the optical system. Throughout this paper, we use terms such as ``PSF substructure'' loosely to refer to any light registered on the detector that originates from the bright central star, regardless of whether it stems from the seeing halo, from the diffraction pattern, or from reflections or scattering within the optical system. The mismatch between the target frames and the reference frames leads to large residuals and inevitably to a poor image depth in the final collapsed image.

With $\alpha$~Pic no longer considered as PSF reference, the only viable option was to use $\beta$~Pic itself for PSF referencing. This can be done using various implementations of angular differential imaging \citep[ADI; see][]{marois2006}, in which several frames of the same target at different rotation states can be used as PSF references for each other. Normally, any observation for which ADI is foreseen would ideally be acquired in so-called pupil-stabilised mode. In pupil-stabilised mode, the telescope pupil is kept fixed on the detector throughout the observation, while the astronomical sky is allowed to rotate relative to the detector as the Earth itself rotates. Because the FORS observations of $\beta$~Pic were performed before any implementations of modern ground-based ADI\footnote{Similar types of processing for space-based telescopes existed at least conceptually for a long term prior to this; see e.g.\  \citet{mueller1987}.} existed and did not foresee any such operations, they were executed in standard field-stabilised mode with the sky fixed on the detector. However, a pupil-stabilised framework can be readily produced during post-processing by simply determining the parallactic angle of each frame, and digitally rotating all frames to a common angle through interpolation. As long as the exposures are not long enough to substantially smear the pupil in the field stabilised observations, this framework is a strong analogy for a set of pupil-stabilised observations. 

The smearing affects all localised static PSF features (e.g.\ static or quasi-static speckles), but in the context of these observations, the most notable features subject to smearing are the diffraction spider arms. Given the 250\,s exposures, some smearing is inevitable; fortunately, the sky placement of $\beta$~Pic means that the rotation rate of the field relative to the telescope pupil is quite uniform so that the smearing is largely the same in all frames. This facilitates reproducing and subtracting in an ADI framework. The issue might have been more serious if the target had resided at a declination closer to -24\degr, where the minimum zenithal angle from the telescope is small. This would have resulted in a highly non-uniform rotation rate, which in turn would produce non-uniform smearing and complicate the PSF subtraction. Hence, while smearing does affect the PSF subtraction quality, it is a manageable effect in the context of our observations. On this topic, we also note that there is in fact a small hidden benefit from using field-stabilised observations for ADI purposes rather than pupil-stabilised observations: In pupil-stabilised imaging, field objects such as planets or disks exhibit smearing to the same extent that PSF features exhibit smearing in field stabilised imaging. Smearing of the field objects is a known complication in conventional ADI \citep[e.g.][]{lafreniere2007} and can be a strong effect for near-zenith targets, potentially reducing the signal-to-noise ratio ($S/N)$ of the planet or disk fluxes. In field-stabilised mode, this effect is thoroughly mitigated. Despite this minor benefit, we assess (in retrospect) that pupil-stabilised mode would have been preferable for these observations, and that it is generally preferable for most ADI applications.

Following the reasoning above, the general procedure for ADI processing of the FORS data is (1) to register the centre and relative parallactic angle of each frame using the spider arms of the the PSF. (2) For each target frame, to choose a subset of the other $\beta$~Pic frames to act as PSF references. (3) To shift and rotate all included frames to a common centre and a common pupil angle. (4) To construct an optimal PSF reference frame from the available PSF representations and subtract it from the target frame. (5) To rotate each PSF-subtracted target frame back to a common field angle and collapse them into a final frame, using a median for the collapse. (6) To filter out any remaining smooth radial residuals from the PSF subtraction by taking the azimuthal median at any given radius from the central star and subtracting it from all pixels at that radius (for all radii). (7) To evaluate any self-subtraction of the $\beta$~Pic disk that may have arisen in the process; substantial self-subtraction of extended flux sources is a common issue in ADI processing. The key steps in this processing are steps 2, 4, and 7, that is,  how to best choose the PSF reference set, how to best construct an idealised master PSF from that set, and how to best characterise the self-subtraction that results as a consequence of this procedure. 

Under favourable circumstances, these choices may have been nearly trivial. In total, the $\beta$~Pic observations span a wide range of parallactic angles, and because the $\beta$~Pic disk is seen nearly edge-on, its morphology is primarily radial, with a relatively limited azimuthal span. Hence, a natural option for any given target frame would be to select almost all other frames for the reference library, removing only those frames with a small (few degrees) relative parallactic angle to the target frame. This is equivalent to imposing a protection angle in conventional ADI-based techniques \citep[e.g.][]{lafreniere2007}, and mitigates or removes self-subtraction by excluding reference frames where the $\beta$~Pic disk would overlap with itself in the target frame. The large number of remaining reference frames can then in principle be used to construct a high-quality optimised reference using some reasonably sophisticated scheme such as Karhunen-Lo{\`e}ve image projection \citep[KLIP, see][]{soummer2012}. 

Unfortunately, this is not a feasible strategy for this set of observations, for several reasons. Firstly, just as the observations are not stable between $\alpha$~Pic and $\beta$~Pic as discussed above, they are not stable either between different observing sessions of $\beta$~Pic itself. The PSF substructure changes from night to night in similar ways as it changes between $\alpha$~Pic and $\beta$~Pic, so that while the contrast comes out better when using only $\beta$~Pic than when using $\alpha$~Pic as reference, it is still very far from ideal. This problem is strongly amplified by the specifics of the opaque bar that was used as a form of coronagraphic mask during all the observations. While modern coronagraphic techniques would typically fix the mask to the pupil of the telescope, the FORS bar is fixed to the field. This has the consequence that when rotating any pair of exposures to a common parallactic orientation, the bar maps over two separate angles in the common image space. Because the bar already takes up a very large fraction of the image space for a single orientation, the combined space it takes up for two different orientations, or for an even larger number of orientations across the full sequence, quickly becomes unmanageable. This is particularly important for an algorithm such as KLIP because if a certain pixel position is behind the bar in any individual reference or target frame, the corresponding pixel position must be masked not only in the frame itself, but in every other frame that is used for the PSF modelling and subtraction. In the end, the usable image space only consists of each pixel that avoids the bar in every single target and reference frame, which for all practical intents and purposes is an empty set if all available reference frames are used. 

\begin{figure}[htb]
\centering
\includegraphics[width=8.5cm]{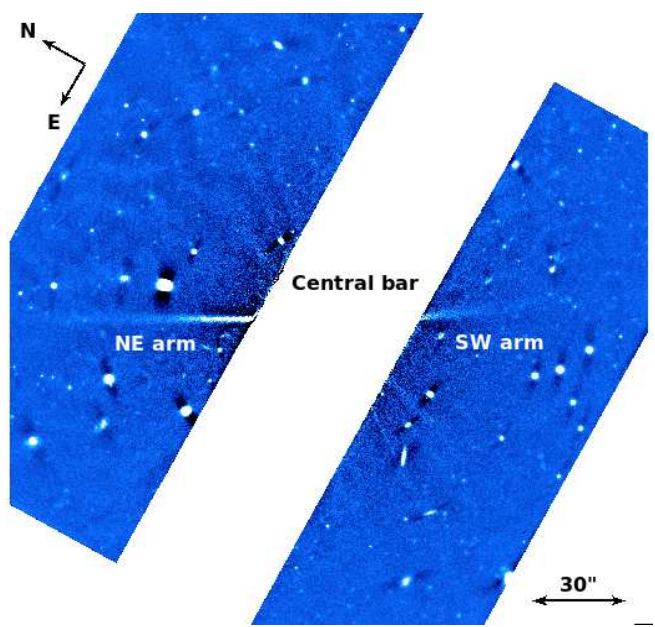}
\caption{ADI-subtracted image of $\beta$~Pic and its circumstellar debris disk. Self-subtraction has inevitably taken place in the procedure, as shown by the dark traces on each side of the disk. Still, the disk flux is detectable out to very large separations.}
\label{f:diskselfsub}
\end{figure}

The only practical solution to these case-specific issues is to instead be very selective in terms of reference frames, and in particular to take advantage of the fact that the stability within observing blocks is much higher than between observing blocks. There are five frames per observing block, so that for each target frame, it is possible to use four reference frames without compromising PSF stability by choosing references from other blocks. This is too few frames for algorithms such as KLIP to perform properly, but a classical median-based PSF construction method \citep{marois2006} works well for the purpose. All PSF-subtracted target frames can then be combined into a final image, regardless of which observing block they originated from. Using this strategy, we reach a much higher image depth and quality in the final image than with any of the other strategies tested and described above. A downside of this strategy is that the field rotation within an observing block is only $\sim$10\degr, so that the effective protection angle is not very large in general. As a result, a rather substantial amount of self-subtraction is imposed in the process (see Fig. \ref{f:diskselfsub}). However, this self-subtraction can be accounted for, as discussed in the next section.

\subsection{Accounting for self-subtraction}
\label{s:selfsub}

A central method for correcting for self-subtraction and derive subtraction-corrected parameters for disks in ADI processing is through forward-modelling with negative disk injection \citep[see e.g.][]{thalmann2014,janson2016,lagrange2016}. In this framework, a parametric model of the disk is constructed and subtracted from each individual frame (at its original orientation) prior to the ADI reduction. The ADI reduction is then performed as normal, and at the end of the procedure, the residuals in the disk region are calculated. If the model matches the actual properties of the disk, the residuals are small, while if the model is a poor match, the residuals are larger. Thus, it is possible to construct a grid of parameter values and run a new ADI reduction for each combination of parameters in the grid. The minimum residual case can be considered as the best parametric fit to the actual data. A perfect fit is fully unaffected by self-subtraction because after subtraction of such a model from the science data, there is no more disk flux that can overlap with itself during the PSF subtraction. We therefore implemented such a scheme on the $\beta$~Pic data in the manner described below.

The parametric disk model can be constructed in several different ways, depending on the implementation. For example, if the disk consists of a single dust belt with an eccentric ring-like shape such as in HR 4796 \citep{milli2017}, a code such as GRATER \citep{augereau1999} can be used to model a simple dust distribution and derive a flux distribution from radiative transfer calculations. In the case of $\beta$~Pic, where the inclination is close to edge-on and we suspect that multiple disk components are overlapping, we chose an approach more like previous work that has been done for $\beta$~Pic at smaller separations by \citet{ahmic2009}, for instance. First, we chose an approximate position angle to represent the disk plane. The two arms of the disk were treated separately with individual angles in this regard because there is a clear angular offset between them, which we return to later. Along the axis defined by the approximate disk midplane, we then characterise the disk as slices of vertical flux distribution parametrised by their width, shape, and amplitude. The analysis was made in a context wherein the image of the disk was rotated such that it was parallel to the horizontal axis of the pixel array (again, separately for the two sides of the disk). Slices of the disk are therefore represented by vertical columns of the pixel array. As a first step, we simply fit Gaussian functions to pixel columns in the self-subtracted ADI image. The free parameters in the fit are the amplitude, width, and centre of the Gaussian, and the minimum $\chi^2$ solution gives an initial estimate for these quantities. These estimates are inevitably skewed due to the self-subtraction, which affects the apparent width and amplitude of the disk slices, and because the real vertical profile differs somewhat from a Gaussian (as we verify in subsequent steps), but it provides a good starting point for defining the grid values of the negative disk-injection procedure that follows as the next step.

Because the negative injection procedure attempts to minimise residuals in the disk region, we need to define a specific region in the final image space in which the residuals are evaluated. For this purpose, we chose a rectangular box centred on the disk midplane that was 6$^{\prime \prime}$ wide in the vertical direction and stretched from separations of 26$^{\prime \prime}$ to 98$^{\prime \prime}$ in the horizontal direction on the brighter NE side, and from 31$^{\prime \prime}$ to 59$^{\prime \prime}$ on the fainter SW side. In addition to the faintness of the SW arm, there is also a group of faint extended objects (probably a clustering of background galaxies) just outside of 59$^{\prime \prime}$, preventing an accurate analysis beyond that point. The inner separations are slightly larger (by $\sim$2$^{\prime \prime}$) than the separation range covered by the central bar, which arises from the fact that the bar edge is non-orthogonal to the disk plane, leaving a thin wedge where a fraction of the disk is visible, but cannot be sampled symmetrically with respect to the disk plane. We simply disregarded this minor part of the image space in our analysis. The vertical profile is characterised by the function $f = p_{\rm a} e^{-\left(| z|/p_{\rm w}\right)^{p_{\rm s}}}$, where $p_{\rm a}$ is the amplitude parameter, $p_{\rm w}$ the width parameter, and $p_{\rm s}$ is the shape parameter, where for example\ $p_{\rm s} = 1.0$ implies an exponential drop-off. In the inner ranges of the $\beta$~Pic disk, the shape has been found in previous studies to be super-exponential, that is,\ with values of $p_{\rm s} < 1$ \citep{ahmic2009}. Before we subtracted the corresponding vertical profile from a given location in the disk, we convolved it with a PSF as derived directly from one of the brighter (but non-saturated) background stars in the field. Becaus an exponential or near-exponential function is very sharply peaked near the midplane, the PSF convolution is an important step even for vertical profiles with a quite large $p_{\rm w}$.

Because different parts of the disk interact azimuthally but not radially in a classical ADI reduction, the optimisation problem of finding a best-fit model becomes separable in the radial direction. In principle, the disk could be divided into radial sections where each section is assigned its own set of parameters prior to the run, but this would be inefficient (or even computationally unmanageable) because every set of parameters needs to be run through its own separate ADI reduction, which is a time-consuming procedure. A much more efficient strategy is to apply parameter grid steps uniformly across the whole disk, except for the output image after each ADI reduction, and then evaluate the residuals locally among the output images. In other words, allowing for different parameter sets to yield the minimum residuals in different radial sections of the disk. In the case of the amplitude, each grid step is a uniform factor that is multiplied with the initial estimation of the radial brightness profile. Because the self-subtraction that we evaluated is expected to vary smoothly with separation, it is not necessary to estimate it in every individual column, and indeed, such a procedure would be strongly affected by random noise in the outer ranges of the disk where the flux is low. Hence, while the innermost 20 pixels of the NE side of the disk are very bright and thus could be easily evaluated individually, for the remaining disk we performed the evaluation over ranges of 100 pixels (approximately 12.5 resolution elements for a seeing of $\sim$0.8$^{\prime \prime}$) at a time on the NE side. The total radial extent of the NE side is approximately 1000 pixels, or $\sim$125 resolution elements, of which 740 pixels are used outside of the central bar. Individual column corrections were then assigned through linear interpolation. The procedure was similar on the SW side, but because the disk is less extended and has no very bright part there, evaluations were performed in steps of 25 pixels across that arm. The procedure was performed in several steps in order to ensure that a global minimum could be reached and to gradually fine-tune the parameters, and initially, this was done only for variations in amplitude and width, keeping the shape fixed at  $p_{\rm s} = 0.83,$ which is what was found for this parameter closer in \citep{ahmic2009}. However, because the intermediate best-fit solution still was not perfect and the residuals visually suggested that the vertical shape was off, we ran an additional procedure in which the shape parameter was allowed to vary along with the width and amplitude. In this iteration, the residuals were evaluated globally across the whole disk region at once. This yielded a best-fit $p_{\rm s} = 1.12 \pm 0.32$, which we use for the remainder of the analysis. Here, the estimated error has been set as the threshold for which the fit residuals in the evaluation region are within 10\% of the minimum residuals. In this context, the slightly wider profile than in the inner disk \citep[0.83, ][]{ahmic2009} is only marginally significant. The $S/N$ per $1^{\prime \prime}$ circular footprint at the disk midplane at the NE side is 30.0 at 650\,AU. At 900\,AU, the disk is approximately two times fainter but the PSF residuals are also smaller, leading to an $S/N$ of 24.9. At the SW side, the $S/N$ is 15.1 at 650\,AU and 6.9 at 900\,AU.

\begin{figure}[htb]
\centering
\includegraphics[width=8.5cm]{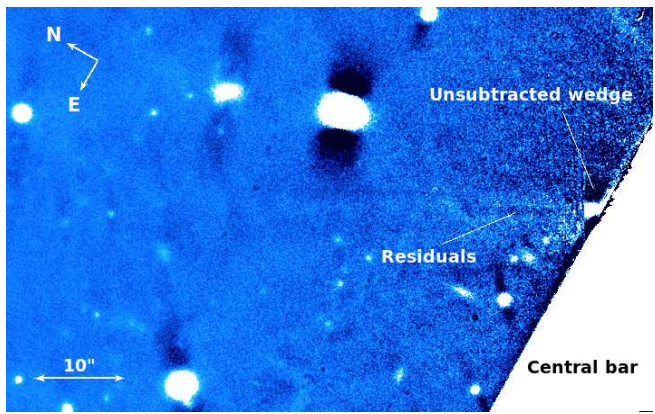}
\caption{Residuals after negative injection of a disk model prior to the ADI processing for the NE arm of the disk. Some residuals remain in the region closest to the star, which may be related to the detailed vertical profile of the disk. However, they are greatly reduced by the fitting procedure.}
\label{f:northeastsub}
\end{figure}

\begin{figure}[htb]
\centering
\includegraphics[width=8.5cm]{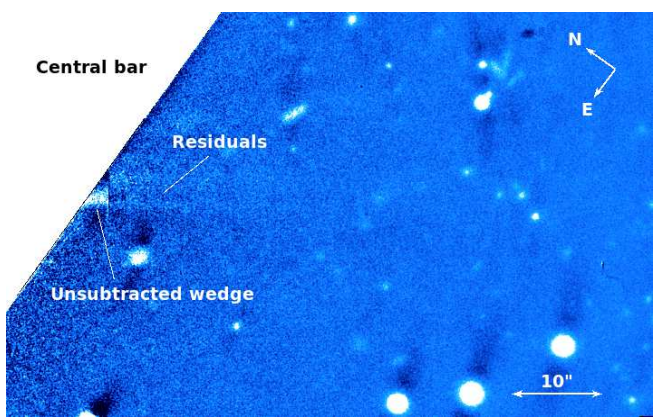}
\caption{Same as Fig. \ref{f:northeastsub}, but for the SW arm of the disk.}
\label{f:southwestsub}
\end{figure}

While some local residuals can still be seen in the resulting images for the NE (see Fig. \ref{f:northeastsub}) and SW (see Fig. \ref{f:southwestsub}) arms of the disk, they are considerably reduced by the fitting procedure. Most likely, the vertical profile is not perfectly represented by the parametric equation that we used, or is a combination of several overlapping components with different parameters. Once provided with the best-fit parameters from the above procedure, we can produce an output image which, to first order, is free from self-subtraction effects. This was done in an iterative process, where the starting point was a tailored ADI procedure in which the best-fit model disk was subtracted from the PSF reference frames, but not from the target frames. The result was an ADI-reduced image in which the disk was visible and free from ADI-induced self-subtraction features. The exact same ADI procedure was then repeated, but in this step, the output image from the previous ADI reduction was used as the input model for the new one. Any negative value in this semi-empirical input model was set to zero because it is an unphysical value. This procedure was then repeated iteratively. With every step the model is expected to be a more accurate representation of the actual disk morphology, and thus subtracting it from the PSF reference frames should yield a yet more accurate output image to be used as a model in the next step. We can assess how well this works by monitoring what happens to the background stars and galaxies in the FORS field of view for which no explicit optimisation has been performed during any step of the process. Successively with increasing iterations, the fidelity of the background sources increased, with decreasing background shadows around them, confirming that the procedure worked as intended. We find that the output results effectively converge after five iterations, where further iterations did not significantly change the output result. Thus, we took the fifth iteration image as our final output, free (to first order) from self-subtraction effects. The latter iterative part of the procedure is similar to the scheme devised by \citet{pairet2018} for the same purpose of achieving accurate disk morphologies in ADI-subtracted data.

In general, this iterative procedure can be expected to work increasingly well at large separations because self-subtraction effects are smaller in this regime to begin with. This means that the template image that is subtracted from the PSF frames in each iteration is more accurate, such that a stronger convergence can be reached faster. At smaller separations, the input image is less accurate, and the convergence is consequently slower or worse. Because physically unrealistic (negative) regions are masked out in the procedure, corrections to the PSF using a non-perfect template should always be as good or better than using no correction at all, however, so that while convergence may be limited in the innermost regions, there should be little or no risk of any divergent solutions. As mentioned above, we are fortunate to have background stars spread around the field, allowing us to check for such effects. In contrast to extended sources such as disks, point sources have a precisely known morphology, so that any effects imposed by the algorithm can be monitored through how they affect the point sources. We find that point sources far out in the field (even very bright ones) lack any discernible subtraction shadows around them in the final output. The closest-in point sources do have weak shadows around them, but are nonetheless well represented by point-source morphologies. This confirms that the procedure does work best at large separations and becomes increasingly challenging at smaller separations, but that even at the smallest separations probed in this study, the algorithm works and improves on the pre-iteration results.

\section{Results and analysis}
\label{s:analysis}

\begin{figure}[htb]
\centering
\includegraphics[width=8.5cm]{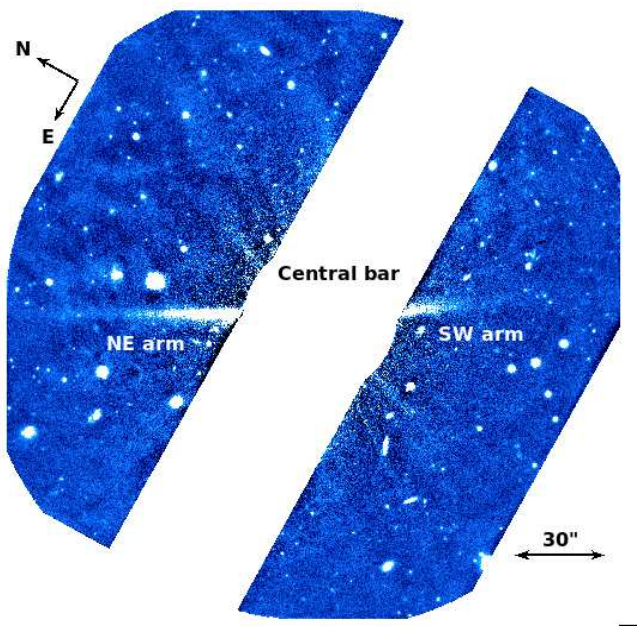}
\caption{Final image after subtracting a best-fit model from the PSF reference frames. The self-subtraction normally imposed by ADI is removed (to first order) by this procedure.}
\label{f:diskcorrected}
\end{figure}

\begin{figure*}[htb]
\centering
\includegraphics[width=16cm]{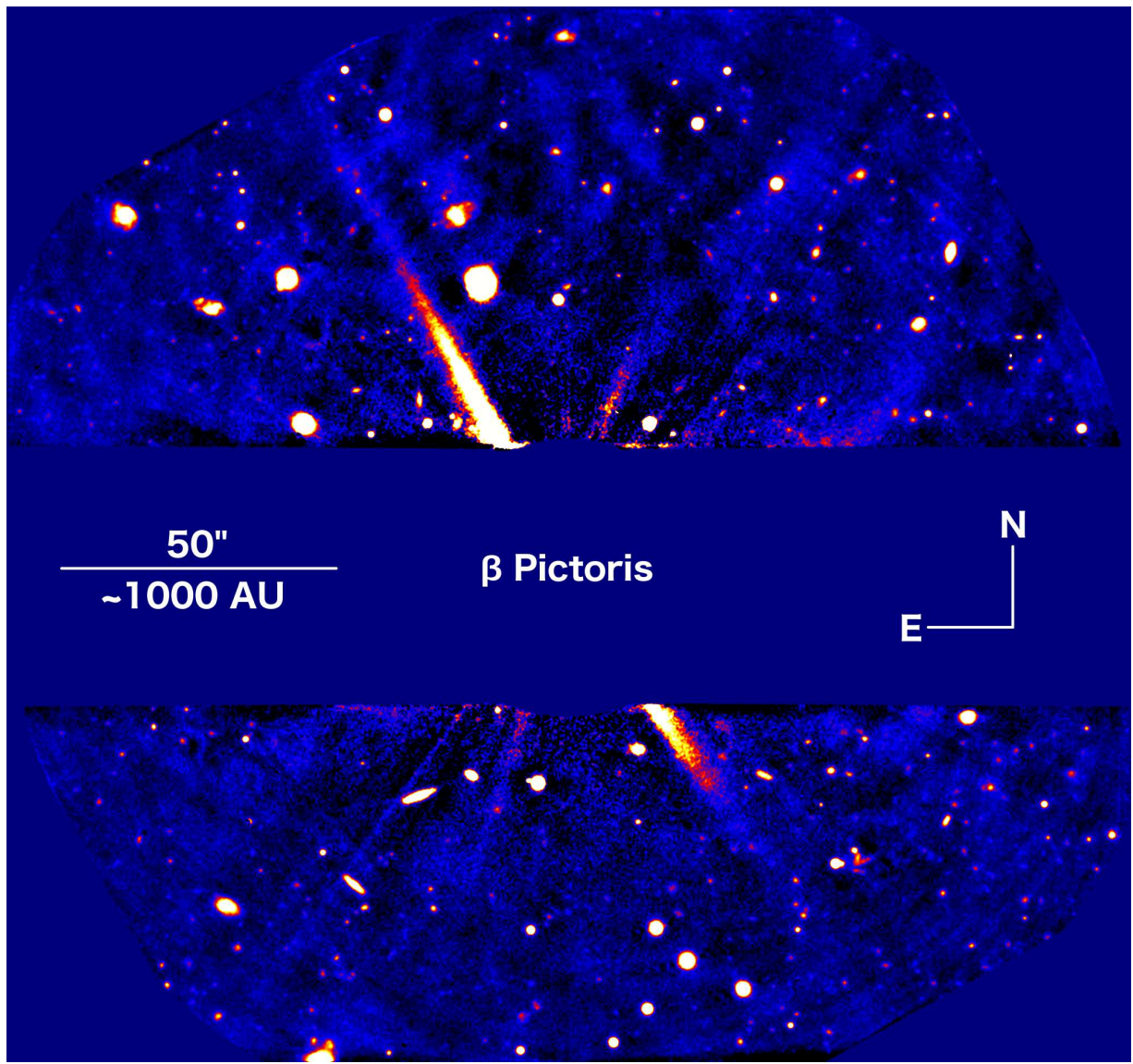}
\caption{Smoothed version of the final image, rotated to the conventional sky orientation with north up and east to the left.}
\label{f:disksmooth}
\end{figure*}

Despite all the challenges related to the PSF subtraction described in the previous section, the final output image shown in Fig. \ref{f:diskcorrected} is of good quality, and constitutes the deepest imaging of the $\beta$~Pic system acquired to date. We also show a version of the image smoothed with a Gaussian kernel with a full width at half maximum (FWHM) of 0.6$^{\prime \prime}$ and rotated to the conventional sky orientation in Fig. \ref{f:disksmooth}. 

We extracted a surface brightness profile along the disk midplane through aperture photometry using a circular aperture with a diameter of 0.6$^{\prime \prime}$, which is close to the FWHM of the seeing-limited PSF. Uncertainties were calculated based on the residuals in a disk-subtracted image, where the iterative disk image was subtracted from both the target and PSF frames, instead of from the PSF frames alone. The result was converted into units of mag/arcsec$^2$ by normalising for the size of the extraction aperture. This conversion also requires a photometric zero-point, which we determined using background stars in the FORS field around $\beta$~Pic. For this purpose, we selected five stars that are all recorded in the \textit{Gaia} DR2 catalogue \citep{brown2018}, bright enough to have a very good significance but not bright enough to be non-linear or saturated in the FORS images, and at a substantial distance from $\beta$~Pic itself in order to minimise the effect of its PSF halo on both the FORS and \textit{Gaia} data. The calibration stars are summarised in Table \ref{t:photref}. Because the FORS data are in V band, a photometric conversion from \textit{Gaia} magnitudes is required, which we acquired using the $G_{\rm BP} - G_{\rm RP}$ colour according to transformation equations given in the ESA online \textit{Gaia} documentation. 

\begin{table*}[htb]
\tiny
\caption{Photometric calibration stars used in the FORS field of view.}
\label{t:photref}
\centering
\begin{tabular}{lcccccccc}
\hline
\hline
Identifier & $G$ & $\sigma_G$ & $G_{\rm BP}$ & $\sigma_{G_{\rm BP}}$  & $G_{\rm RP}$ & $\sigma_{G_{\rm RP}}$ & $V$ & $\sigma_V$ \\
  & (mag) & (mag) & (mag) & (mag) & (mag) & (mag) & (mag) & (mag) \\
\hline
Gaia DR2 4792773281421225600    &       18.9501 &       0.0066  &       19.0672 &       0.0473  &       17.7094 &       0.0148  &       19.296  &       0.025   \\
Gaia DR2 4792773285714718848    &       19.0758 &       0.0063  &       19.1299 &       0.0610  &       18.0628 &       0.0294  &       19.298  &       0.026   \\
Gaia DR2 4792773285714718976    &       19.0135 &       0.0064  &       18.8574 &       0.0389  &       18.1578 &       0.0257  &       19.121  &       0.013   \\
Gaia DR2 4792774694465890688    &       16.6359 &       0.0017  &       17.5452 &       0.0158  &       15.5409 &       0.0024  &       17.363  &       0.011   \\
Gaia DR2 4792774728825628160    &       17.0194 &       0.0016  &       18.0984 &       0.0274  &       15.8351 &       0.0058  &       17.940  &       0.022   \\
\hline
\end{tabular}
\end{table*}

\begin{figure}[htb]
\centering
\includegraphics[width=8.5cm]{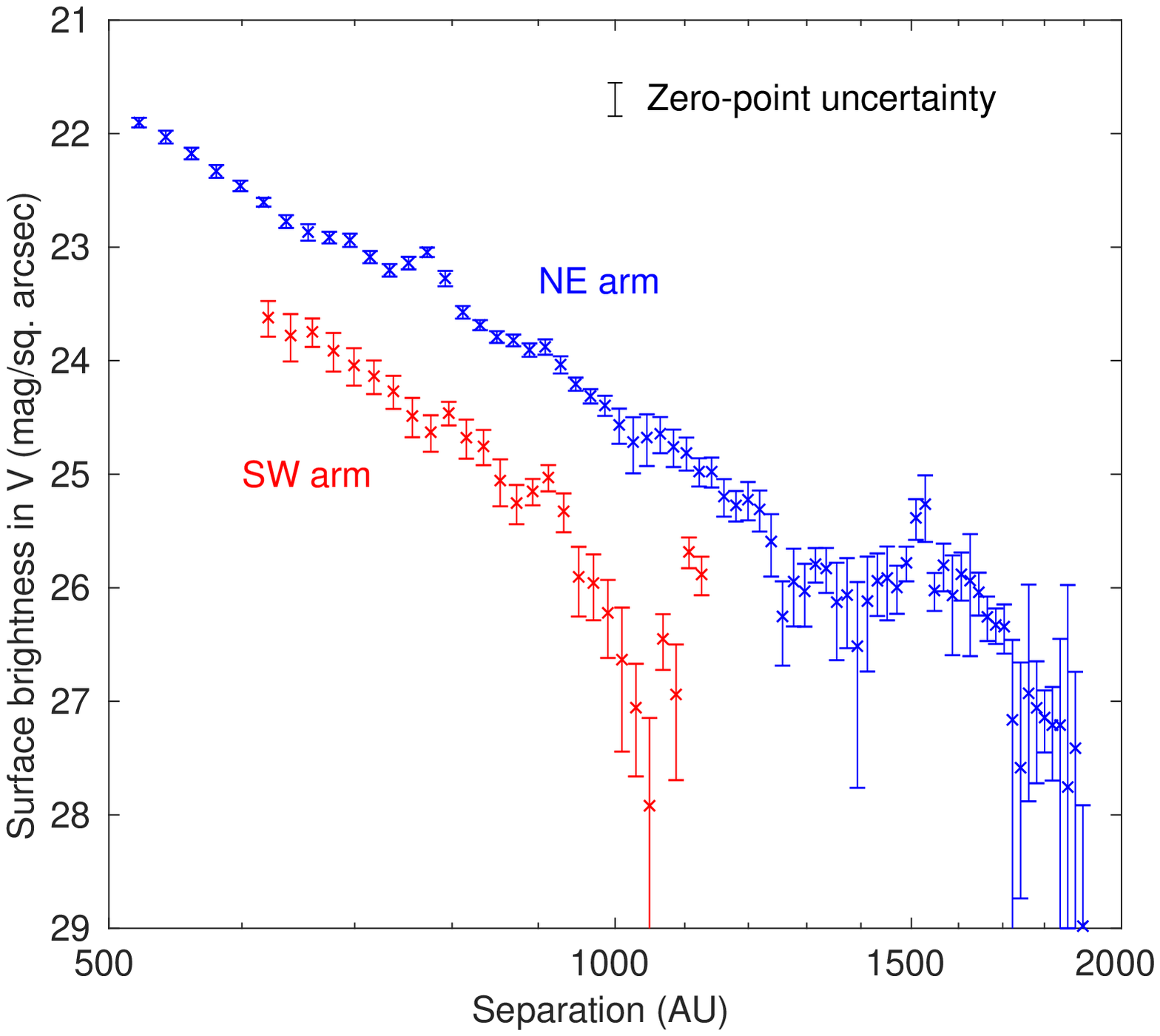}
\caption{Surface brightness profile for the brighter NE arm (blue) and the fainter SW arm (red) of the disk. The bumps that deviate from the average trends are probably caused by background galaxies (see text). The SW arm measurements start at a larger projected separation than for the NE arm because of a small asymmetry in the placement of the star behind the central bar.}
\label{f:armsbright}
\end{figure}

In Fig. \ref{f:armsbright} we show the resulting surface brightness profile in calibrated photometric units. The uncertainty in the photometric calibration is dominated by the scatter among the calibration stars and is 0.14 mag. This scatter might arise from the effect of the bright stellar halo on the \textit{Gaia} data. PSF effects around bright targets are known to have substantial effect on stars around them in DR2 \citep[e.g.][]{brandeker2019}; this will most likely improve in subsequent \textit{Gaia} releases. Because the zero-point calibration uncertainty affects all brightness profile data points systematically in the same way, we plot it separately in Fig. \ref{f:armsbright} rather than adding it in quadrature to the intrinsic photometric uncertainties there. The calibration clearly dominates the uncertainty in the absolute level in the bright inner parts of the disk (particularly in the NE arm), while intrinsic uncertainties dominate in the faint outer parts. In the outer parts of the FORS field of view, the median 3$\sigma$ sensitivity (per PSF footprint) is 27.7 mag in $V$ band. The previous deepest image of the $\beta$~Pic disk at any comparable wavelength was presented in \citet{larwood2001}, who achieved an R-band 3$\sigma$ sensitivity of 25.8 mag at these separations. Thus, the high depth of the FORS image makes it a valuable resource for characterising the outer ranges of the $\beta$~Pic disk.

In addition to the general smooth drop-off in brightness with increasing distance from the star, there is apparent substructure in the form of bumps at specific radial locations. This substructure in the $\beta$~Pic disk has been noted before in the inner parts of the disk \citep[e.g.]{kalas2000,wahhaj2003}. One possible interpretation of these bumps is that they represent an ansa from a discrete ring in the edge-on disk, but another possibility that we need to account for, in the context of the outer ranges of the $\beta$~Pic disk, is an overlap between the disk and a background galaxy. In these ranges, we are probing very faint flux at considerable image depth, which means that we are at the border of being confusion limited with respect to galaxies in the background. Many background galaxies can indeed be seen across the FORS field, therefore it would not be surprising if some of them overlapped or partially overlapped with the area covered by the disk. 

One bump that stands out in the surface brightness profiles in Fig. \ref{f:armsbright} is located in the NE arm of the disk between 750 and 800\,AU. This feature has in fact been noted before as feature `A' in \citet{kalas2000}. Because almost a decade of baselines has passed between the original image of feature A from 1995 and the FORS image from 2004, we can distinguish between a disk feature and a background feature on the basis of whether it shares a common proper motion with the star and disk. We find that the photocentre of the bump resides at a separation of $\sim$39.7$^{\prime \prime}$ in 2004, while in 1995, it resided at a separation of $\sim$40.7$^{\prime \prime}$. While astrometric centring on the bump is very difficult due to its overlap with the disk, the precision is certainly much better than the FWHM of the PSF of 0.6$^{\prime \prime}$. This means that the bump has certainly moved with respect to the star. Moreover, the NE proper motion of $\beta$~Pic of $\mu_{\rm RA} = 4.65$ mas/yr and $\mu_{\rm Dec} = 83.10$ mas/yr \citep{vanleeuwen2007} means that a static background object would move in a relatively radial direction toward the star, and by a similar amount over a decade as observed for bump A. We thus conclude that the most probable interpretation of bump A is a background galaxy. If this interpretation is correct, future imaging of the bump should place it at an even closer separation to the star, and it should gradually start separating from the NE arm on the southern side of it. Another conspicuous bump in the FORS data occurs at the outer edge of where we measure the SW arm. As we noted previously, the outer edge of our SW arm measurements is set by the fact that a group of faint objects overlap with the extended direction of the arm. These objects appear to be extended and may represent a clustering of galaxies in the background. The measured bump might mark the first such galaxy, and we thus consider it unlikely that this is a feature within the disk itself. Future imaging for relative proper motion measurements is required to fully establish the nature of this bump because it is smoothly extended and quite faint and therefore difficult to visually distinguish from disk flux in the images. The outermost range (beyond $\sim$1400\,AU) in the NE arm also seems to indicate some broad bump, which is subject to the same uncertainties as the outer bump on the SW side.

Because of the suspected (and in one case verified) background nature of the blobs discussed above, we rejected the corresponding areas when making power-law fits to the general flux slopes of the arms. This was made through line fitting to the flux distribution versus separation in logarithmic space, defining $\alpha$ as the power-law index on the projected radial dependence $r^{\alpha}$. In the NE arm within 1200\,AU we find that $\alpha_{\rm NE} = -3.7 \pm 0.1$, and when we included the whole range out to nearly 2000\,AU, the value remained very similar at $\alpha_{\rm NE} = -3.9 \pm 0.2$. In the SW arm out to 900\,AU, the slope is similar or very slightly steeper at $\alpha_{\rm SW} = -4.0 \pm 0.3$, but then becomes significantly steeper such that the average slope of the full arm out to 1050\,AU (where the outer bump starts) is $\alpha_{\rm SW} = -6.0 \pm 0.8$. If interpreted as a power-law break around 900\,AU, the outer slope out to 1050\,AU is extremely steep at $\alpha_{\rm SW} = -18.0$ and would imply a sharp outer cut-off for the SW arm, but we note that this latter value is based only on a small section in one of the faintest parts of the disk. The general trends are consistent with the fact that the brightness asymmetry between the NE and SW side, which is clearly present also in the inner parts of the disk \citep[e.g.][]{kalas1995}, is even more extreme in the outer parts that we probe here. The slopes on either side out to $\sim$1000\,AU are very similar to the corresponding slopes at $\sim$100--300\,AU calculated by \citet{kalas1995} as $-3.76 \pm 0.05$ on the NE side and $-4.07 \pm 0.05$ on the SW side.

\begin{figure}[htb]
\centering
\includegraphics[width=8.5cm]{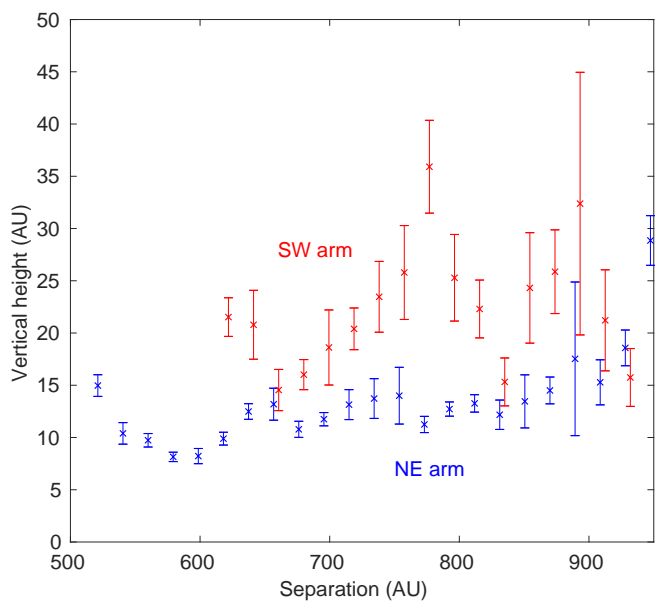}
\caption{Vertical height of the brighter NE arm (blue) and the fainter SW arm (red). The SW arm appears to be slightly broader on average at these separations in terms of its scale height above the midplane.}
\label{f:armsheight}
\end{figure}

The vertical height and trace of the disk can also be derived from the ADI-corrected PSF-subtracted image, but it is a more challenging prospect than for the surface brightness profile because the brightness profile is measured in the disk midplane where the flux is highest, but the height and trace depend strongly on the behaviour in the wings extending vertically from the midplane, where the flux is lower and more susceptible to the confusion limit mentioned above. We therefore restricted this analysis to within separations of 50$^{\prime \prime}$. We performed an additional round of fitting for the width and trace, where we fit vertical profiles with the $p_{\rm s} = 1.12$ exponent determined in section \ref{s:selfsub}, and with the surface brightness profile in the midplane fixed as determined above. In this round of fitting, we downsampled the data by averaging groups of 10-pixel columns and individually fitted the height and centroid of each averaged column. The best-fit heights are shown in Fig. \ref{f:armsheight}. The SW arm appears to have a generally broader profile at these separations, with a mean height of $22.3 \pm 1.4$\,AU compared to the mean height of the NE arm of $13.4 \pm 0.9$\,AU.

\begin{figure}[htb]
\centering
\includegraphics[width=8.5cm]{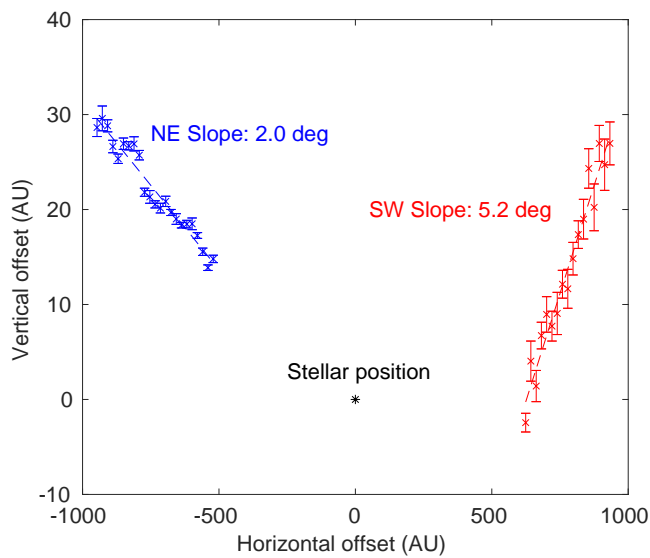}
\caption{Disk trace of the brighter NE arm (blue) and the fainter SW arm (red). The disk has a clear tilt, with both arms deviating from the average disk plane at smaller separations, which is the reference plane in this figure. This implies a high degree of forward-scattering and/or a significant deviation from an edge-on inclination (see text). The spatial scalings of the x- and y-axes are very different, which causes a strong visual exaggeration of the tilt angles.}
\label{f:armsnomirror}
\end{figure}

\begin{figure}[htb]
\centering
\includegraphics[width=8.5cm]{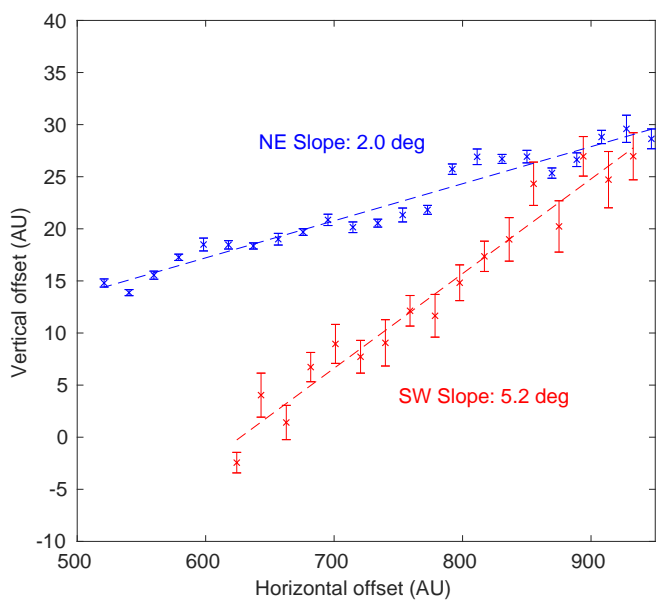}
\caption{Same as Fig. \ref{f:armsnomirror}, but with the absolute value of the horizontal offset (i.e. mirroring the NE arm) in order to enhance visibility.}
\label{f:armstrace}
\end{figure}

The disk traces of each arm are shown in Fig. \ref{f:armsnomirror} with an alternative display in Fig. \ref{f:armstrace}. They confirm the tilt angle between the arms that we noted previously. We made linear fits to the traces and related them to the average position angle of the inner disk of $\sim$30.8 deg \citep[e.g.][]{kalas1995}. The NE side has a differential angle of 2.0 deg in a clockwise (CW) direction in the image plane with respect to the reference angle, while the SW side has a differential angle of 5.2 deg in a counter-clockwise (CCW) direction. The total tilt angle between the arms is therefore 7.2 deg. This is higher than in the inner parts of the disk \citep[e.g.\ 1.3 deg in][]{kalas1995}. A tilt angle between two arms of a nearly edge-on disk arises naturally if there is strong forward-scattering in the disk and if the disk has a non-zero angular offset from being perfectly edge-on. In this context, because the tilt angles are asymmetric relative to the inner disk, it would require an angular offset between the inner and outer disk planes not only in inclination, but also in the disk position angles. The higher the degree of forward-scattering and the larger the offset angle, the higher the expected tilt angle. 

We can thus use the tilt angle to constrain the inclination and scattering anisotropy, but with some degeneracy between them. To do this, we generated simple models with different inclinations and degrees of forward-scattering in the GRATER code \citep{augereau1999}. In GRATER, the forward-scattering can be quantified by the Henyey-Greenstein index $g$, which has a value between -1 (maximally back-scattering case) and +1 (maximally forward-scattering case). We input values of 0.2 to 0.9 in steps of 0.1 for $g$ and inclinations $i$ of 85 to 89 deg in steps of 1 deg. In addition, because the tilt angle can in principle depend on what the location is relative to parent planetesimal belt that the dust is expected to originate from, we tried three different values of the belt radius $r_0$: 400, 500, and 600\,AU. The dust density in the model drops off slowly from the ring centre, consistent with the trend observed in the surface brightness profile; in this case represented by a power-law index in the density of -1.8. For each simulated disk, we evaluated the tilt angle between the two arms of the disk by fitting Gaussian centroids to sequential vertical slices of the disk between 30$^{\prime \prime}$\ and 50$^{\prime \prime}$\ ($\sim$600--1000\,AU), and then we performed a linear fit to the resulting centroids to find the average slope of the arm relative to the main disk plane. The model disk is circular and single-component, and is not meant to represent the actual morphology of the disk (which is much more complex), but only to provide a handle on what is required to produce tilt angles of the size observed.

The dependence of the tilt angle on $r_0$ is found to be generally weak in this context, but the dependence on both $i$ and $g$ is strong. Interestingly, we find that both a higher degree of forward-scattering and a larger offset from edge-on (i.e. smaller inclination) is required for the outer disk than has previously been determined for the inner primary disk. A $g < 0.8$ or an $i > 86$ would both have great difficulty to reproduce the tilt angle. A combination of $i = 85$--86 deg and $g = 0.8$--0.9 is necessary to acquire a tilt angle of 7.2 deg. While $i = 85$--86 deg is consistent with the inclination range determined in \citet{kalas1995}, it is lower than the $i = 87.7$ deg determined for the inner primary component of the disk in \citet{ahmic2009}. The $g$ in the inner range is determined as 0.3--0.5 in \citet{kalas1995} and as 0.64 and 0.85 for the primary and secondary components of the disk, respectively, in \citet{ahmic2009}. A value of $g = 0.8$--0.9 for the outer disk is similar to the secondary disk in the inner range, but higher than the primary disk. $g = 0.9$ would be consistent with the degree of forward-scattering for dust, for example, in comet McNaught \citep[C/2006 P1; see][]{marcus2007}. 

The difference in forward-scattering could therefore be explained by a gradient in the dust properties in the disk. The inclination might, for instance, imply a warp in the disk, where the outer disk has a systematically different orientation than the inner disk, much like how there are two inner disk components with systematically different orientations already known in the system. The asymmetry in the tilt angle of the two arms that we measured with respect to the inner disk plane would further support this scenario. Because a considerable brightness asymmetry between the arms exists and probably implies at least some strongly eccentric subcomponent of the disk, this might also contribute to the asymmetry in the tilt angle: The two arms might then contain relative populations that are asymmetrically distributed outside of the image plane, with one population of dust primarily in front of the star at one side and another population primarily behind the star at the other, affecting their relative scattering properties even if the underlying population of dust is the same.

\section{Discussion}
\label{s:discussion}

As we have noted above, there are multiple asymmetries in the $\beta$~Pic disk that become increasingly extreme farther out in the disk. Asymmetries are present not only in the dust, but also in the second-generation gas that has long been known to exist in the disk \citep{slettebak1975,olofsson2001,brandeker2004}. The dissociation timescales of the detected molecules imply that the gas may have been released in a series of events following a disruption of a large planetesimal in the recent past \citep{cataldi2018}. The disruption might have occurred as a result of a collision with another planetesimal \citep{jackson2014,lawler2015} or through tidal disruption in a close encounter with a massive planet nested within the disk \citep{cataldi2018,janson2020}. A  large number of fragments of different sizes would have been formed in the disruption, and acquired a range of eccentric orbits with different orbital elements, but which all converge once per orbit to the disruption location where they originally formed. Over time, this creates a series of additional collisions at this location, which can maintain a localised release of short-lived gases over $\text{several }$thousand years (several orbital timescales). The observed gas in the disk is predominantly found in the SW arm at a projected separation of $\sim$50--100\,AU, which marks the origin of the event and the mutual periastron in this scenario. Meanwhile, the fragments and associated dust would be spending most of their time at apastron, at substantially larger projected separations, at the other side (i.e. the NE arm) of the disk. It might therefore be the case that the excess flux associated with dust that we observe in the NE arm could in part be related to the recent disruption event. In this framework, it might be viewed as one highly eccentric component of the disk superimposed on a less eccentric, more ordered outer disk around $\beta$~Pic, much like how there are (at least) two distinct components in the inner disk.

Despite $\text{about }$17 years of technical developments, it is a far from trivial prospect to improve the image depth and wide-separation contrast at visible wavelengths acquired in the observations presented here. As mentioned in Sect. \ref{s:obs}, the observational programme spans $\text{about }$7.5 hours of VLT time, of which 2.8 hours are effective integration time on $\beta$~Pic, under good seeing conditions and in a dark lunar phase. The disk is probed down to a surface brightness of 27--28 mag/arcsec$^2$. The sky brightness \citep[$\sim$22 mag/arcsec$^2$ at Paranal during the time of observation, see][]{patat2003} is obviously an important factor in this regard, but still, the large size of the VLT and high sensitivity of FORS gives it a significant advantage even with respect to space-based facilities. Acquiring a 27 mag/arcsec$^2$ sensitivity in a 5$\times$5 pixel box at 5$\sigma$ significance with the WFC3 camera at the Hubble space telescope, for example, would require 39 hours of effective on-source integration time, and the field of view of 162$^{\prime \prime}$ would be insufficient to cover the whole disk in one shot. From the ground, PSF-limited visible light imaging has benefited from adaptive optics developments \citep[e.g.][]{schmid2018,close2018}, but because the isoplanatic angle is small, the good performance reached within a few arcseconds of separation cannot be extended to the image scales required to cover the $\beta$~Pic disk. Nonetheless, in possible future imaging efforts for probing the outer disk, some clear improvements might be made based on our experiences with the FORS data set. Most importantly, the bar used to mask the star in the images was fixed to the sky rather than to the pupil. This made any highly sophisticated PSF subtraction schemes essentially impossible. If an observational setting could be used in which the mask would be fixed to the pupil, and ideally, in which the detector would also be pupil stabilised, this would greatly enhance the achievable contrast performance. Observations should be acquired in long uninterrupted sequences because the largest PSF variations were experienced between nights (or between target and reference star); this procedure would also allow us to accumulate a substantial amount of parallactic angle variation across the observation, which further benefits contrast in ADI-related applications. Observations in different photometric bands than $V$ may be particularly useful because the colour of the dust might help constrain dust sizes and similar characteristics. A first indication of dust colour might be provided by the fact that the $\beta$~Pic disk appears to be at least marginally visible out to $\sim$1900\,AU in the \citet{larwood2001} $R$-band image, even though that image is $\sim$2 mag less deep than the FORS image. This could imply a redder colour in the scattered disk flux than for the central star, which has a typical A-type colour of $V - R = 0.12$ mag. Polarimetric imaging could be an alternative or complementary characterisation pathway, although its success heavily relies on a reasonable high degree of polarisation for the dust involved. 

Having been probed from single-AU distances in the form of transiting comets \citep[e.g.][]{vidal1994,zieba2019} out to $\sim$2000\,AU in the FORS images presented here, the observational data for the $\beta$~Pic debris disk cover a truly vast range, and provide information on its characteristics at distances from the star that are comparable to the inner Oort cloud in the solar system; a unique feature in the known debris disks. We also note that radar measurements tracking micrometeorites that enter the Earth's atmosphere have led to the detection of a population of particles whose kinematics imply an interstellar origin \citep{baggaley2000}. Trace-back studies have pointed out $\beta$~Pic as the most probable origin of this population \citep{krivov2004}. It might therefore be argued in this sense that the observable range of the $\beta$~Pic disk even reaches out to scales of tens of parsecs.

\section{Conclusions}
\label{s:summary}

We have examined a deep FORS archival imaging data set of $\beta$~Pic in order to determine the fundamental properties of the outer ranges of its debris disk. One of the main challenges in this context is the complex and varying PSF structure of the bright primary, particularly because these observations were acquired before ADI became widely applicable. This meant that special considerations had to be taken in the subtraction procedure, and that the subtraction had to be performed over data sets spanning a rather small range of parallactic angles. As a consequence, substantial self-subtraction was inevitable as a consequence of the procedure, and in response to this, a dedicated negative injection procedure had to be implemented to mitigate these effects. 

Aside from the impressive scale spanned by the far-reaching arms of the disk, its most prominent characteristic is the extreme asymmetry between the NE and SW arms of the disk. The asymmetry exists already in the inner disk region, but it becomes increasingly enhanced with increasing separation, and manifests itself both in the surface brightness, in the vertical height, and in the trace of the disk with respect to the midplane. One apparent component of this asymmetry includes discrete bumps in the surface brightness profiles, but we note that at the faint brightnesses that are being probed, the observations are in a partially confusion-limited regime with respect to background galaxies, and we thus conclude that the most likely underlying cause of the bumps is blending with background galaxies. In the case of the most prominent bump of the NE arm, we confirmed this hypothesis by relating its position to previous epoch imaging and noting that its motion resembles that of a static background object. The asymmetries in the total brightness, brightness slope, height, and trace are however all real physical effects within the disk, and reflect asymmetries in the morphology of the underlying dust. Eccentric subpopulations of dust are a likely cause for many of the observed feature, some of which might be related to the localised gas emission previously reported in the SW arm of the disk, which in turn has been theorised to be caused by a recent large disruption event within the disk.

\begin{acknowledgements}
M.J. gratefully acknowledges funding from the Knut and Alice Wallenberg Foundation. This study made use of the CDS services SIMBAD and VizieR, and the SAO/NASA ADS service. We thank Matthias Samland for useful discussion, and the anonymous referee for the careful review and useful comments that helped improve the manuscript.
\end{acknowledgements}

\end{document}